Manuscript

# Cruise Missile Target Trajectory Movement Prediction based on Optimal 3D Kalman Filter with Firefly Algorithm

*Mahdi Mir\**,
*Department of Electrical Engineering, Ferdowsi University of Mashhad, Mashhad, Iran*

**Abstract**

It is hoped that there will never be a war in the world, but one of the defensive requirements of any country during the war is the missiles used for destruction and defense. Todays, missiles movement from origin to destination is an important problem due to abundant application of missiles in wars. This is important because of the range of some missiles is low and other are very high. Parametric indeterminacy are several factors in missile movement prediction and trajectory such as speed, movement angle, accuracy, movement time, and situation and direct control. So this research trying to provide a method based on LQR controller with 3d Kalman filter and then set motion and specify path without deviations based on Firefly Algorithm. It is expected that the results of an appropriate evaluation can be obtained by simulating the MATLAB environment and graphic display of a cruise missile.

**Keywords**: Cruz Missile, Missile Movement Prediction and Trajectory, LQR Controller, 3d Kalman Filter, Firefly Algorithm.

## 1- Introduction

Missile routing scheduling is an efficient method for preventing probabilistic errors for collission. In recent years, routing scheduling algorithms had a lot of attention and used in various applications. Routing scheduling algorithms separated in two groups based on variuos search method such as decisive and random search [1]. Decisive search methods include dynamic programming such as A star method [2], but evolutonary algorithms used in random search methods such as Genetic and PSO algorithm [3].

Optimal route collect information directly by a connected line by using greedy and evolutionary algorithms in a set of road station in 3d space. So the initial route of these algorithms produce as a broken line. However, a missile can't determine accurately schedule with a broken line and it's because of dynamic performance and missile kinematic features limitation. The main porpuse of routing tracking is producing a distinct path which the radius of curvature is greater than the minimum radius of rocket rotation at any point and the path should identify the continuity of the curve [3].

---

\* Corresponding Author.
E-mail address: Mahdimir.ir@gmail.com (Mahdi Mir)

In general, tracking the missile involves identifying missile targets from launch to target time. Meanwhile, the identification of movements parameters such as speed, angle of movement, accuracy, time of movement, positioning and positioning of the head and the bottom of the rocket as direct movement is important, which is the main challenges of this research. Hence, a large study area for military and aeronautical applications is being made to track missiles and other moving objects. Typically, the dimensions first obtained to track the objectives of a moving rocket by a sensor in the polar coordinates of the rocket are reported, then modeled by Cartesian coordinates. For this job, Kalman filters are appropriate. The Kalman filter is a filter that can detect noise as a variable, estimate errors and possible errors, and also estimate unknown variables that tend to be accurate. . To do this, there are several Kalman filter models, including the Linear Kalman Filter[1], the Extended Kalman Filter[2], the Without Sequence Kalman Filter[3], the Particle Kalman Filter[4], and the 3D Kalman Filter[5].

In tracking the path, assuming that the rocket parameters change, and this change is done linearly, Kalman filter relationships can be overcome to this problem. Because the simulation world (the continous world) is different from the real world (the discrete world), the same parameters should be used to implement the proposed method that tracks the missile until it encounters the target, and between track-to-target tracking it will be done. Hence, the Kalman filter which has more similarities to the discrete world is a three-dimensional Kalman filter that is considered in this study. Because the environment in which the missile moves is 3D, and the Kalmni filter, which can track target operations by specifying parameters such as speed, angle of movement, precision, movement time, position, is Three-dimensional Kalman filter which has higher capacities than other Kalman filters, including the Extended Kalman filter and the Without Sequencing Kalman filter.

It should be noted at the outset that the parameters of the missile, including its position, speed, and initial motion angle are set manually and for which purpose it is determined and these parameters change until reaching that goal. Output noise data is the sensor which sends the information to the controller. In fact, there is also a controller in the rocket that can be controlled by any model. The purpose of this study is to use the LQR controller.
One of the reasons for using the 3D Kalman filter is as follows: tracking the path, assuming that the parameters of the rocket are changed, and this change is done linearly, so that Kalman filters can be used to overcome these problems. The three-dimensional Kalman filter parameters to reach the target in this research include a number of important items which include the position of the rocket, the rocket's velocity at run-time, the time-per-second, and time-varying velocity update at different time intervals along the movement, rocket control to track the target, and prevent collisions with other objects (barrier detection). These parameters cannot be fully and correctly optimized with Kalman filters. Now why are we considering space as NP-Hard, because to this day the correct and accurate approach has not been scientifically presented in the articles, and therefore the use of evolutionary algorithms and swarm intelligence can be optimized for this space.

---

[1] LKF
[2] EKF
[3] WSKF
[4] PKF
[5] 3DKF

The method of timing and tracking the path to reaching the intended purpose of this research is to use random search methods. Since the Genetic algorithm has high convergence and speed, but lower family algorithms that is swarm intelligence, have a higher rate of improvement than evolutionary algorithms. Hence, the use of a Firefly Algorithm will be used to optimize the 3D Kalman filter. The Firefly Algorithm is very effective in places with a very large search space compared to places with small search space and is very effective in optimizing it due to its non-singular characteristics. The two most important factors in this algorithm include the attractiveness and intensity of light. Methods for reaching the target state estimation to describe the dynamic state of the missile to reach the target will be considered including equivalent noise, detection and input along with switching. Considering the time, speed, movement and changing model, is vital that according to the definitions given by Kalman filters and its Three-Dimensional type by optimizing it in a large space with the Firefly Algorithm, these challenges are fixed.

In the following, in addition to considering the method outlined in [3], it is necessary to consider some other similar methods that will be considered as base articles. In [4], the selection of the self-adaptive parameter is used with a prediction approach using a Hidden Markov Model[6] algorithm for a moving object. Also, in [5], dynamical analysis and path tracking are used by calculating the torque method optimally for a moving robot. In [6], the Kalman filter has been developed to detect rockets that consider parameters such as position and velocity to reach the target as critical research parameters that have been performed well.

## 2- Literature Review

Due to the increasing spread of space science and communications, many researchers have examined the problem of automatic tracking of the target from different aspects. The radar tracking system generally determines the direction of the missile towards the target based on the energy emitted from the targets. These systems are often confronted with a real target diagnosis in the face of misleading fighter action in destroying information that has an impact on the production of guidance commands. In general, routing in missiles can be divided into flat routing, hierarchical routing, and routing of a network-dependent location. This research uses a routing method based on network structure.

Each missile needs a controller to predict the movement. It is necessary to control a system, collect information, process it and issue appropriate commands to the operating units that drive the system [7]. The type of controller of a missile is comparative in that there are generally three types of comparative controllers that include the benefit tabulation, comparative control of the reference model and self-regulating regulators [7]. This research uses self-regulating regulators.

Machine Learning methods are used to carry out routing operations and to predict cruise missile movement and tracking. Learning the machine as one of the most extensive and widely used artificial intelligence branches, arranges and explores the methods and algorithms by which computers and systems can learn; computer programs can over time, improve their performance based on received data [8]. One of the methods that are considered in the routing and tracking of missiles which are part of the machine learning family are evolutionary algorithms and subsets of these swarm intelligence algorithms. Swarm intelligence is a systematic property in which the agents collaborate locally and the collective behavior of the entire agent leads to a convergence at a point close to the optimal global answer. The strength of these algorithms is the absence of a global control [9-11]. The method used by this research as the swarm intelligence algorithm to predict the movement and to achieve the main goals of the research is the Firefly Algorithm.

---

[6] HMM

## 3- Proposed Approach

The prediction of rocket motion and tracking is a nonlinear optimization problem. The main purpose of tracking the missile is to move correctly without dealing with obstacles in the path and adjusting the control variables to optimize one or more objective functions, while at the same time, a series of tracking constraints such as route finding and route shortening. The missile tracking problem is mathematically formulated in the form of Eq. (1).

Min Func(x, u)

Subjected to: $h(x, u) \leq 0$

and $g(x, u) = 0$ (1)

According to Eq. (1), Func is the objective function that needs to be optimized, h is the inequality constraint that represents the limitations of the missile tracking agent, g is the constraints equal and is represented by the non-linear equation $g(x, u)$ and x is the vector of independent variables or state variables, and u u the vector of control variables or independent variables. The control variables including the output manufacturer include:

- ✓ $P_G$ is the real power of routing,
- ✓ $P_{G1}$ is the simple route tracking mode,
- ✓ $V_G$ is the missile manufacturer's voltage in tracking,
- ✓ $T_S$ is setting the missile transmission route,
- ✓ $Q_C$ is the missile shunt compensator is in tracking.

Due to the control variables in the tracking of the missile, there is Eq. (2).

$$U^T = [P_{G2} \ldots P_{G_{NG}}, V_{G1} \ldots V_{G_{NG}}, T_{s1} \ldots T_{s_{NT}}, Q_{C1} \ldots Q_{C_{NC}}]$$ (2)

According to Eq. (2), NG, NT, and NC, respectively include path generator numbers, missile routing set numbers, and missile power compensation numbers. In this research, the longitudinal motion of the rocket is investigated to track routing operations. The missile math model is also needed to allow other operations to be performed. The equation of longitudinal motion of a rocket can be represented by Eq. (3) for linear motion.

$$\begin{pmatrix} \dot{u} \\ \dot{w} \\ \dot{q} \\ \dot{\theta} \\ \dot{h} \end{pmatrix} = \begin{pmatrix} X_u & X_w & X_q & -g\cos(\theta) & 0 \\ Z_u & Z_w & Z_q & -g\sin(\theta) & 0 \\ M_u & M_w & M_q & 0 & 0 \\ 0 & 0 & 1 & 0 & 0 \\ -\sin(\theta) & -\cos(\theta) & 0 & 1 & 0 \end{pmatrix} \begin{pmatrix} u \\ w \\ q \\ \theta \\ h \end{pmatrix} + \begin{bmatrix} X_{\delta e} & X_{\delta t} \\ Z_{\delta e} & 0 \\ M_{\delta e} & 0 \\ 0 & 0 \\ 0 & 0 \end{bmatrix} \begin{bmatrix} \delta_e \\ \delta_t \end{bmatrix}$$ (3)

Which according to Eq. (3), $t = u\sin(\theta) + w\cos(\theta)$, $\delta_e$ and $\delta_t$, are elevators and control inputs of the rocket, u is the forward speed (horizontal) of the rocket, w is the rocket's vertical speed, q is the ground and ground rates for the initial rocket propulsion, θ is the ground angle and h is the rocket's height to the ground, and $X_u$, $X_w$, $X_q$, $Z_u$, $Z_w$, $Z_q$, $M_u$, $M_w$ and $M_q$ as well as $X_{\delta e}$, $X_{\delta t}$, $Z_{\delta e}$ and $M_{\delta e}$ are the later derivatives missile stability. The mathematical model in Eq. (3) can be represented by the form of Eq. (4).

$$\dot{x} = Ax + Bu \tag{4}$$

According to Eq. (4), $x^T = [u \ w \ q \ \theta \ h]$ is the vector of the longitudinal motion of the rocket, and the Eq. (5), is the transition matrix of the rocket, as well as the Eq. (6), the distribution of control with the 3D Kalman filter, and $u^T = [\delta_e \ \delta_t]$, is the input vector of the control.

$$A = \begin{bmatrix} X_u & X_w & X_q & -g\cos(\theta) & 0 \\ Z_u & Z_w & Z_q & -g\sin(\theta) & 0 \\ M_u & M_w & M_q & 0 & 0 \\ 0 & 0 & 1 & 0 & 0 \\ -\sin(\theta) & -\cos(\theta) & 0 & 1 & 0 \end{bmatrix} \tag{5}$$

$$B = \begin{bmatrix} X_{\delta e} & X_{\delta t} \\ Z_{\delta e} & 0 \\ M_{\delta e} & 0 \\ 0 & 0 \\ 0 & 0 \end{bmatrix} \tag{6}$$

Now, we need to analyze the stability of the rocket model. The longitudinal equations can be calculated using the transfer function, which is the Eq. (7) in theory.

$$(s^2 + 15.043s + 78.0719)(s^2 + 0.587s + 1.1174) = 0 \tag{7}$$

In this research, LQR controller is considered as the optimal control method in the missile. Considering a system with a state space model according to Eq.(8), the vector of optimal control is as follows.

$$u(t) = -Kx(t) \tag{8}$$

In order to determine the optimal control inputs when optimizing state variables at instant, the cost function, also known as the quadratic performance index function, is used to minimize it by using Eq. (9).

$$J = \frac{1}{2}\int(x^T Q x + u^T R u)dt \tag{9}$$

Which according to Eq. (9), Q is a semi-positive matrix of definite symmetric and R is a definite symmetric positive matrix. Matrices of weight Q and R are chosen to control any effective control based on efficiency. The Gain vector matrix of optimal control is calculated as Eq. (10).

$$K = T^{-1}(T^T)^{-1}B^T P = R^{-1}B^T P \tag{10}$$

Therefore, based on Eq. (10), the optimal control equation is converted into Eq. (11).

$$A^T P + PA - PBR^{-1}B^T P + Q = 0 \tag{11}$$

If a definite P positive matrix can be calculated by the rational equation, then the missile will be in the form of Eq. (12) stable movement.

$$A^T P + PA - PBR^{-1}B^T P + Q = 0 \tag{12}$$

Based on a series of specific conditions, the LQR controller can be designed using a rational equation such as Eq. (11) that can be used without changing the control matrix and mode to find the optimal Gain matrix. The optimized 3D Kalman filter can also be determined by considering that there are several unspecified modes. For a rocket without adding an integer, we can use the previously defined matrix A and B to control it by means of a control, and the matrices of the weight of the mode Q and R are used to find the optimal matrix or K, which allows the input of the control gives $u = -kx(t)$.

In this research, the LQR controller was designed without affecting measurement impediments in the estimation and tracking of missile targets. The equations of the system have been discontinued using Euler's approach. Initially, the LQR controller was evaluated without affecting the disturbances, and then the system response to the controller under conditions of disturbance would be investigated, once using the Kalman filter, three-dimensional, and once without using it. Also, operations for estimating and tracking the missile target will be carried out during routing time and reach. A three-dimensional Kalman filter will use state equations (state space matrices) and initial values for calculating gain and residual values, as well as estimating the actual value of the signal in the estimation and target tracking. The steps of the 3D Kalman filter are investigated using linear discrete states and measurement equations, which are in the form of Eq. (13) and (14).

$$X(k+1) = AX(k) + Bu(k) + Gw(k) \tag{13}$$

$$y(k) = HX(k) + v(k) \tag{14}$$

According to the equation of state in Eq. (13), X(k) is the rocket mode vector, A is the rocket propulsion matrix, u(k) is the rocket input vector, B is the control distribution matrix, w(k) is the random Gaussian noise vector with zero mean and covariance structure, G is the missile noise transmission matrix, which is Gaussian noise. In the equation of measurement, according to Eq. (14), y(k) is the measurement vector, H is the measurement matrix, v(k) is the measured noise vector with mean zero and the covariance structure. There is no relationship between the noise of the rocket w(k) and the measurement noise v(k). The covariance matrices for w(k) and v(k) are also calculated as Eq. (15) and (16) respectively.

$$E[w(k)w^T(j)] = Q(k)\delta(kj) \tag{15}$$

$$E[v(k)v^T(j)] = R(k)\delta(kj) \tag{16}$$

According to the above two equations, E is the expected value and $\delta(kj)$ is the Kronecker symbol. The optimal Three-Dimensional Kalman filter, which estimates the rocket mode vector, is carried out with the returning equations that follow. The extrapolation equation is calculated by Eq. (17).

$$X_e\left(\frac{k}{k}-1\right) = AX_e\left(\frac{k-1}{k-1}\right) + BK_{LQR}(k-1)(X_d - X_e\left(\frac{k-1}{k-1}\right)) \tag{17}$$

This sequence is shown in general and summary in Eq. (18).

$$\Delta(k) = Z(k) - HX_e\left(\frac{k}{k}-1\right) \tag{18}$$

Also estimates the state calculated by Eq. (19).

$$X_e\left(\frac{k}{k}\right) = X_e\left(\frac{k}{k}-1\right) + K(k)\Delta(k) \tag{19}$$

The Gain matrix of the optimal Three-Dimensional Kalman filter is also calculated from Eq. (20).

$$K(k) = P\left(\frac{k}{k}\right)H^T R^{-1}(k) = P\left(\frac{k}{k}-1\right)H^T\left(HP\left(\frac{k}{k}-1\right)H^T + R(k)\right)^{-1} \tag{20}$$

The covariance matrix of the Three-Dimensional Kalman filter error is also calculated by Eq. (21).

$$P\left(\frac{k}{k}\right) = (I - K(k)H)P\left(\frac{k}{k}-1\right) \tag{21}$$

The covariance matrix of the extrapolation error is also calculated by Eq. (22).

$$P\left(\frac{k}{k}-1\right) = AP\left(\frac{k-1}{k-1}\right)A^T + BD_u(k-1)B^T + GQ(k-1_G^T) \quad (22)$$

According to Eq. (18) to (22), $X_d$ is the arbitrary vector, and I is the identification matrix. The 3D Kalman filter attempts to estimate the actual signal amount in the target tracking and its estimation with missile-based disturbances based on the firefly algorithm that is carried out through the Gaussian distribution. In the future, improvement of the prediction of the motion and trajectory of the cruise missile route is done using the Firefly Algorithm.

## 4- Simulation

Given the fact that various parameters for different parts of the cruise missile are mentioned in this study, at the beginning of the simulation, the set values should be noted. Inputs in this control include speed, type of longitudinal or deep motion, torque and route identification, and output is path detection and target tracking. In the beginning, different parameters of a cruise missile can be found in Table (1).

**Table (1) Cruise missile parameters**

| Cruise missile rotational capability | $\pi/180$ |
|---|---|
| Gravity(m/s$^2$) | 9.81 |
| Frequency (Hrz) | 60 |
| Light Speed(m/s) | 299792458 |
| Boltzman Constant | 17 |
| Sampling rate for simulation (specific for Euler integral) - (sec) | 0.5 |
| Fly Time (sec) | 270 |
| Environment Size (Km) | 12x12 |
| Cruise missile cruiser height (m) | 200 |
| Average speed at flight time (m/s) | 30 |
| Minimum speed at flight time (m/s) | 15 |
| Maximum speed at flight time (m/s) | 50 |
| Spin speed in radius (m/s) | 200 |
| Initial cruise missile gain | 1 |
| Path to goal (m) | 2500 |
| Internal engine heat engine cruise missile (centigrade) | 23 |

Similarly, the parameters of the Firefly Algorithm are according to Table (2).

**Table (2) Firefly Algorithm Parameters**

| Iteration Number | 1000 |
|---|---|
| Initial Population of Fireflies | 1 |
| Light intensity rate | 0.3 |
| Attraction rate | 2 |

When the simulation is performed, the longitudinal motion of the cruise missile first shows its output. Fig. 1, divided into four sections, shows the response of each step of the cruise missile movement from the onset of flight from the ground to the steady, stable level of flight in the sky.

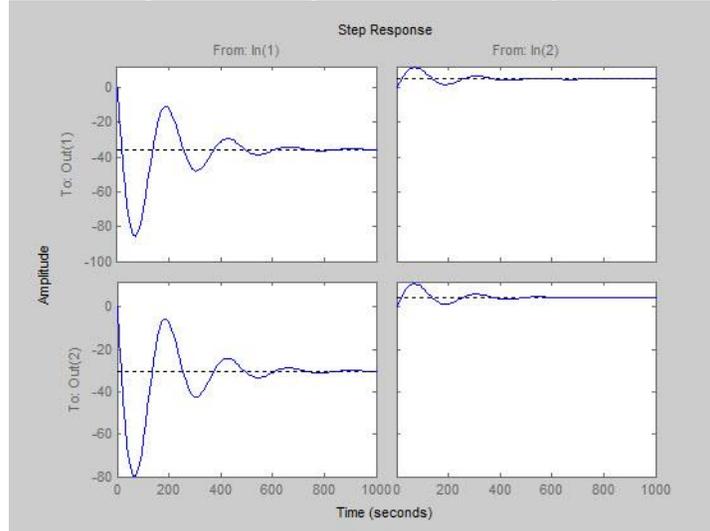

**Fig. (1) Shows the cruise missile output from the beginning of the flight to the equilibrium point in flight in the sky**

Depending on the Fig. (1) and the settings made for the cruise missile, it is shown how the cruise missile moves at any moment. In all the domain images, from the time of flight to reaching a specific point, is shown. On the upper left, the cruise missile range is shown initially, which is intended to move from the ground. In the upper left-hand figure, the range of motion is shown from the surface of the ground, reaching a given point of flight. In the lower left-hand figure, the range of balance adjustment is shown on the fly. In the bottom right-hand side, the equilibrium point is shown. Based on this range of motion adjustment, four Fig. (2) to (5) are also shown for setting conditions.

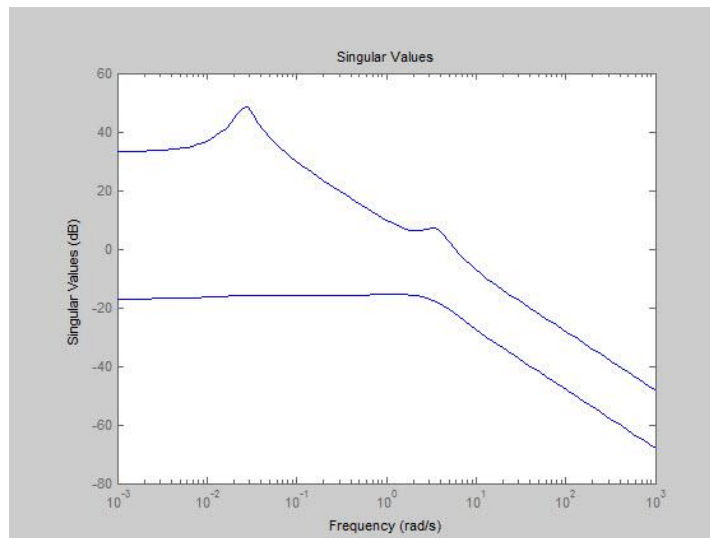

**Fig. (2) Cruise missile range with the intention of moving from the ground**

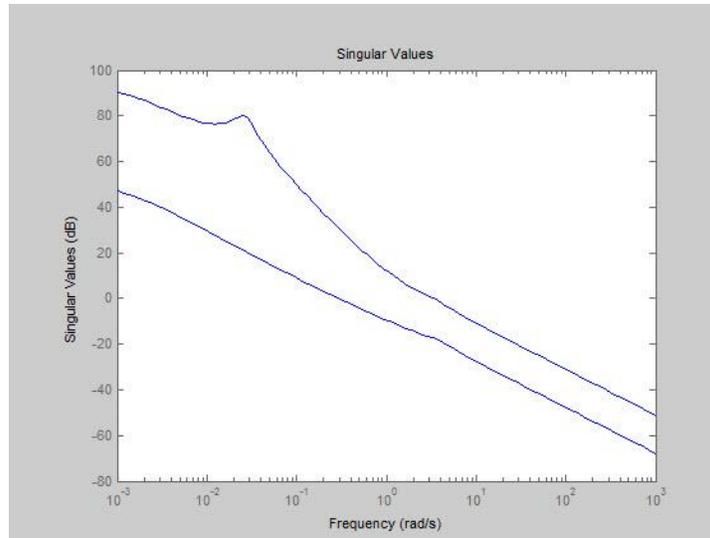

**Fig. (3) The range of motion from the ground to a specific point of flight**

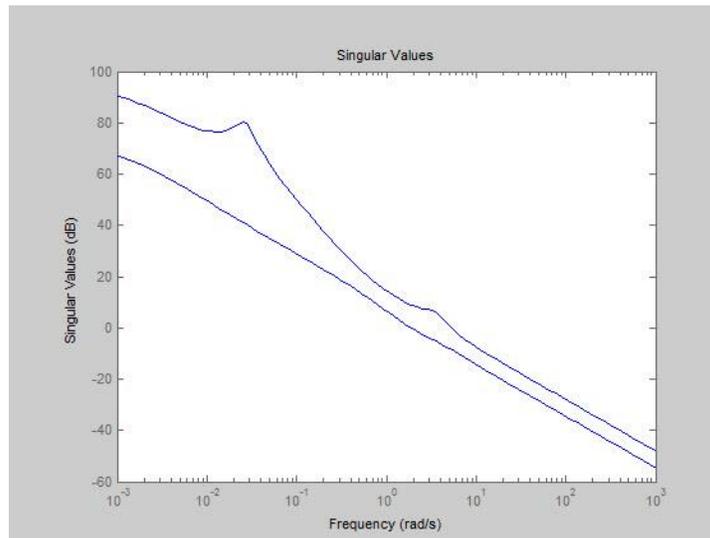

**Fig. (4) Flight adjustment range**

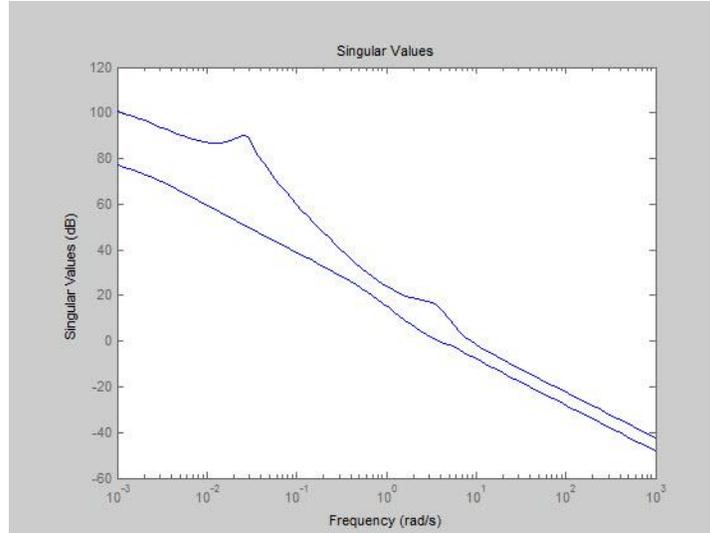

**Fig. (5) Reaching the range of the equilibrium point**

The stable mode is also shown in Fig. (6) after applying the motion. This section will be based on the LQR controller. Given this shape, it can be seen that the cruise missile is in perfect balance and the controller provided in the cruise missile is resistant to the settings. According to Fig. (6), it can be seen that the rate of delay in moving cruise missiles is commensurate with the flight time to reach the target, and is declining.

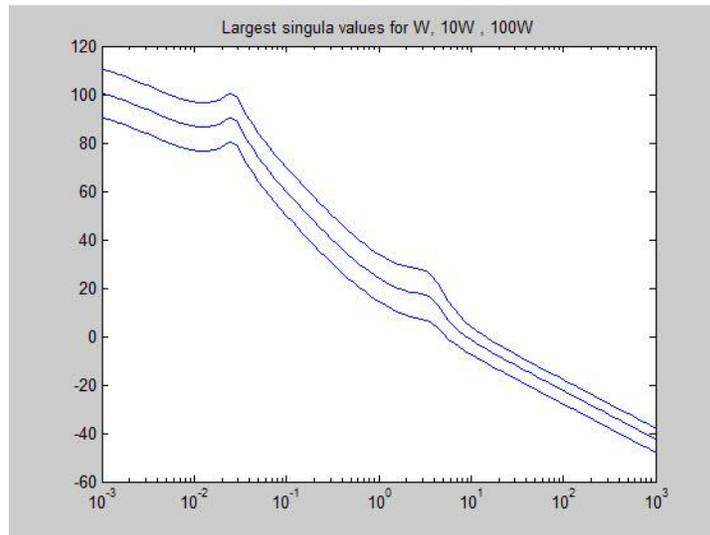

**Fig. (6) Rate of delay in cruise missile movement**

After the operation, the cruise missile output will be shown. The graphical form of it for moving in an environment of a given path is based on a LQR controller based on a Three-Dimensional Kalman filter based on the Firefly Algorithm and the property that has all evolutionary algorithms and swarm intelligence being randomly randomized and may be in each run time from a path, move and start moving to reach the target. But the prediction of moving and tracing it with this algorithm is constant and tracks it until it reaches the target. In Fig. (7), we can see the cruise missile's graphical output.

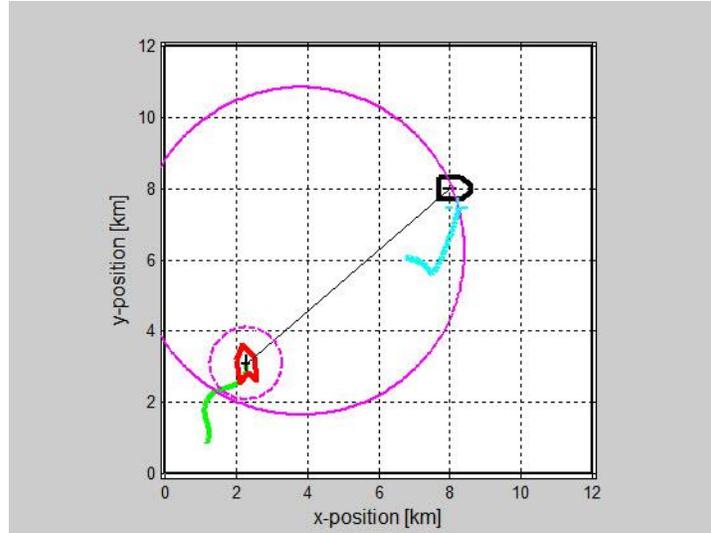

**Fig. (7) The initial movement of the cruise missile at a specified path and tracing it to reach the target**

According to Fig. (7), the cruise missile will move in a 12x12 km environment. The D spot, which is specified at 8x8 km coordinates, is the main objective that the cruise missile must take to reach its path. This is a path traversal using the routing, modeling and its outputs are shown in Fig. (1) and below, from Fig. (2) to Fig. (5), and eventually reaching a state balance was flying and resistant in Fig. (6). The cruise missile is shown in red color. The green lines are the cruise missile route. There are two pink circular lines, one spotted near the cruise missile, which detects the motion of a cruise missile to track the target or D, and tracks the cruise missile. Also, the large pink circle performs route estimation according to the direction the cruise missile runs. Initially, this is a large circle and, to reach the target, it will be as small as possible until the estimated operation is done correctly. Several illustrations of the cruise missile outputs to reach the target are shown in Fig. (8) to (11).

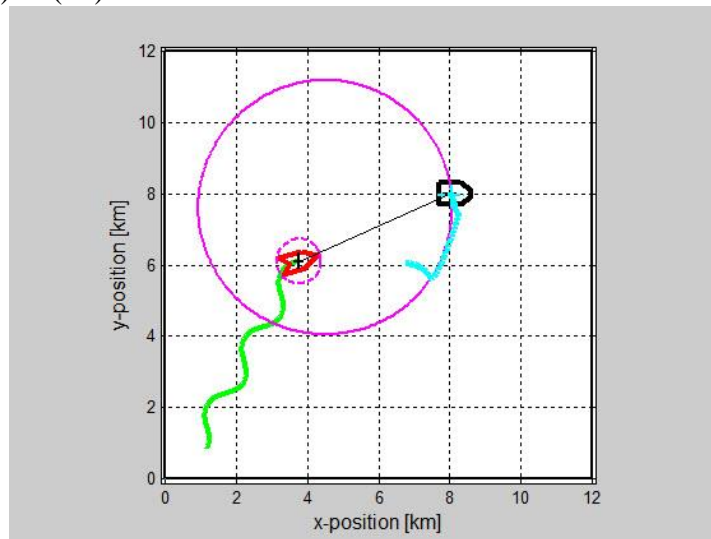

**Fig. (8) Shows the movement of the cruise missile from the green path and tracing and estimating the target**

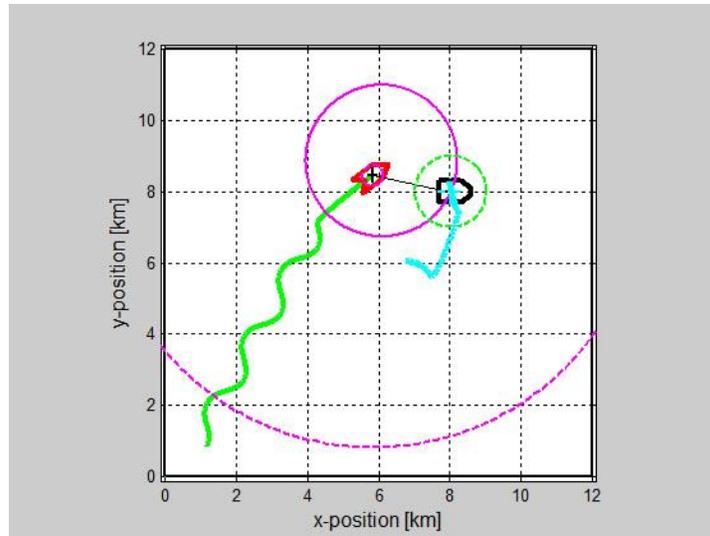

**Fig. (9) Shows the movement of the cruise missile from the green path and tracking and target estimation - approaching the target or D**

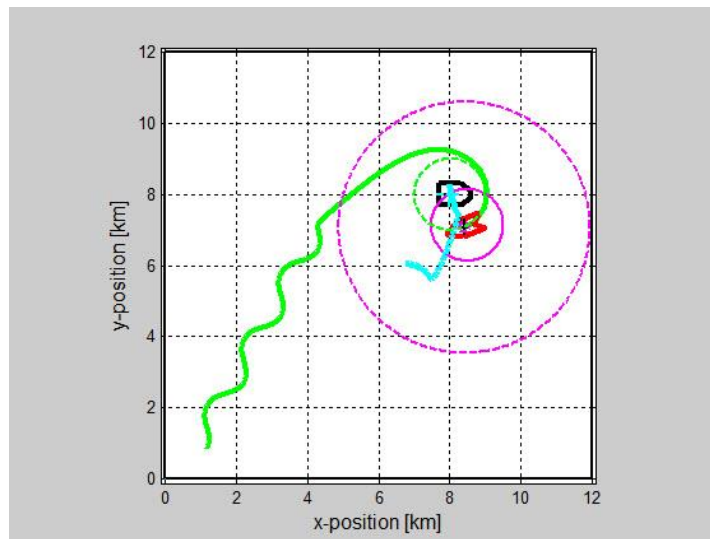

**Fig. (10) Shows the movement of the cruise missile from the green path and tracing and objective estimation - crossing the green circle of the dots around the target or D**

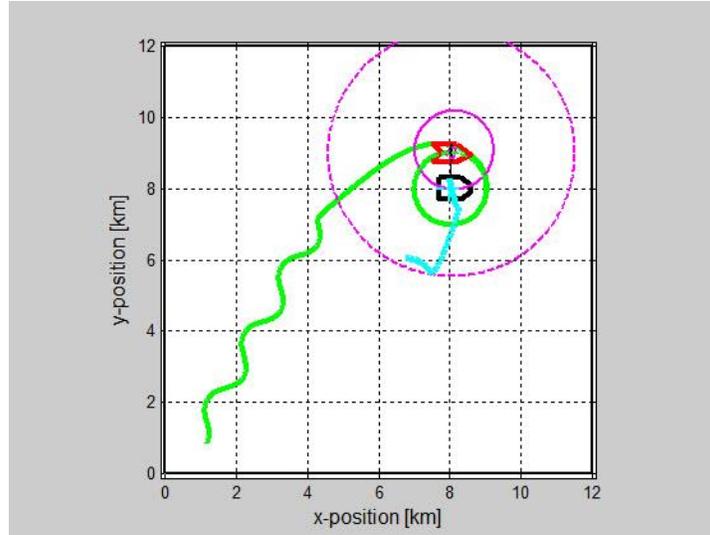

**Fig. (11) Finishing the simulation and passing a specific path to the goal with its tracking and estimating**

Upon completion of the simulation, the cruise missile power consumption is equal to 56.2045 mW, which indicates a good passage. In this research, several evaluation criteria will be used including Mean Square Error[7], Peak Signal to Noise Ratio[8], Signal-to-Noise Ratio[9], and Accuracy criteria. Based on the results of the evaluation criteria after the project implementation, it can be ensured that the proposed method is used to roam the cruise missile and estimate and track the target. In Table (3), the results of the evaluation methods are shown.

**Table (3) Results of proposed approach in terms of evaluation criteria**

| Specificity (%) | Sensitivity (%) | Accuracy (%) | SNR (dB) | PSNR (dB) | MSE |
|---|---|---|---|---|---|
| 80.07 | 80.08 | 96.00 | 56.0618 | 9.9310 | 0.6400 |

## 5- Conclusion

The missile tracking is known as an important issue in military science. Today, the design of missiles is tracked to reach and reach targets, and are capable of controlling and redirecting. The proposed approach of this study is to use evolved methods to track missiles until the goal is achieved. For this purpose, after modeling the cruise missile system and positioning and deploying it, the specified path and the same path are proposed as an optimization space. The rocket tracking route uses a combination of 3D Kalman filter combinations and a Firefly Algorithm. In the future plans, we try to make a simulation of the proposed approach and prove it and compare it with the basic articles of this research.

---

[7] MSE
[8] PSNR
[9] SNR